\documentclass[aps,twocolumn,floats,nofootinbib,superscriptaddress,showpacs,amssymb,amsmath]{revtex4}
\usepackage{graphicx}
\usepackage{bm}
\usepackage{epsfig,subfigure}

\begin{document}
\newcounter{TC}
\newcommand{\useTC}[1]{\refstepcounter{TC}\label{#1}\arabic{TC}}
\newcommand{\MeVcsq}{${\rm MeV}/c^2$}
\newcommand{\GeVc}{${\rm GeV}/c$}
\newcommand{\EGEV}{${\rm GeV}/c$}
\newcommand{\PGEV}{${\rm GeV}^2/c^2$}
\newcommand{\MGEV}{${\rm GeV}/c^2$}

\title{Indications of the possible observation of the lowest-lying $1^{-+}$ QCD state}

\author{Kwei-Chou~Yang}
\affiliation{Department of Physics, Chung-Yuan Christian University,
Chung-Li, Taiwan 320, R.O.C.}

\date{March, 17, 2007}

\begin{abstract}
We discuss properties of  $1^{-+}$ exotic meson within the
framework of the QCD field-theoretic approach. We estimate the mass
of the lowest-lying $1^{-+}$ exotic meson using
renormalization-improved QCD sum rules, and find that the mass lies
around $1.26\pm 0.15$~GeV, in good agreement with the $\pi_1(1400)$
data. This state should be expected in QCD. We find that the mass
for the lowest-lying strange $1^{-+}$ meson is
$1.31\pm 0.19$~GeV. Our result hints that the $K^*(1410)$ may be the
lowest-lying $1^{-+}$ nonet state.
\end{abstract}


\maketitle


\section{Introduction}

Quantum Chromodynamics (QCD) is the presently accepted theory of
strong interactions among quarks and gluons. Perturbative QCD has
been developed in great detail and tested successfully. However, the
physics of the nonperturbative QCD seems to be a much more difficult
task. The rapid development of the lattice QCD theory may offer some
answers to the nonperturbative dynamics.

In addition to normal mesons ($\bar q q $) and baryons ($qqq$),
which can be built up from the naive quark model, QCD allows the
existence of exotic states which can be glueballs, hybrid mesons
(the bound states of $\bar q q g$), tetraquark mesons (the bound
states of $\bar q q \bar q q $), and other multi-parton states.
These exotic states, which go beyond the description of the naive
quark model, can offer the direct evidence concerning confining
properties of QCD.

In the past years possible evidences have accumulated for the
existence of $1^{-+}$ exotic states \cite{PDG}. Now we have two
$1^{-+}$ exotic states below 2 GeV. A negatively charged exotic
state, $\pi_1 (1400)$, with $J^{PC}=1^{-+}$ was observed in $\pi^- p
\to \eta \pi^- p$ \cite{thompson97,chung99} and in $\bar p n \to
\pi^0 \pi^-\eta$ \cite{cbarrel_1}. The corresponding neutral state
was reported by the Crystal Barrel \cite{cbarrel_2} and E852
\cite{dzierba} collaborations in the reactions of $\bar pp\to \pi^0
\pi^0 \eta$ and $\pi^- p \to \eta \pi^0 n$, respectively, where the
decay channel $\eta \pi$ is isovector and hence cannot be confused
with a glueball. Unlike the charged $\eta \pi^-$ channel, the charge
conjugate is a good quantum number in the neutral $\eta \pi^0$
system, where the $\eta$ was detected in its $2\gamma$ decay mode
\cite{dzierba} and very recently in its $\pi^+\pi^-\pi^0$ decay mode
in the E852 experiment \cite{Adams:2006sa}. The advantage of
detecting the $\eta\to \pi^+\pi^-\pi^0$ mode over the all-neutral
final state is that the decaying vertex is determined by charged
tracks. The mass of the neutral exotic $1^{-+}$ state observed in
the very recent E852 experiment is $1257\pm20\pm25$~MeV
\cite{Adams:2006sa}, which is lower than the mass $1360\pm25$~MeV in
the Crystal Barrel measurement. The world average is $1376\pm
19$~MeV \cite{PDG}.

Another observed $1^{-+}$ exotic state, $\pi_1 (1600)$, has also
attracted much attention. In contrast to $\pi_1 (1400)$, the
$\eta^{\prime} \pi$ coupling of this state is unexpectedly stronger
than the $\eta \pi$ coupling although the phase space favors the
latter. Due to the above property, it was argued that the $\pi_1
(1400)$ may be favored for a four-quark state, while the $\pi_1
(1600)$ may be a hybrid meson \cite{Close:1987aw}.

Theoretically, the compositions of observed $1^{-+}$ exotic states
remain unclear. The flux-tube model \cite{Close:1987aw} and lattice
calculations \cite{Bernard:1997ib} predict the lowest-lying $1^{-+}$
hybrid meson to have a mass of about 1.9~GeV. Moreover, the former
predicts that the $1^{-+}$ hybrid meson are dominated by $b_1 \pi$
and $f_1\pi$ decays in contrast with the $\pi_1 (1600)$ experimental
results where the three final states $b_1\pi, \eta^\prime \pi$ and
$\rho\pi$ are of comparable strength, $1: 1.0\pm0.3: 1.5 \pm 0.5$
\cite{Amsler:2004ps}.

To evaluate the mass of the lowest-lying $1^{-+}$
hybrid meson in the QCD sum rule approach, in the literature, people used
\begin{equation*} J_\mu(x) = i \bar d(x)
\gamma_\alpha g_s G_\mu^{\ \alpha}(x) u(x)\,,
\end{equation*}
as the relevant interpolating current in the calculation of the
two-point correlation function. The earliest QCD sum rule
calculations were done in Refs.~\cite{Govaerts:1984bk,Balitsky:1986hf}, where the radiative corrections
and anomalous dimensions of $J_\mu$ and some operators in
operator-product-expansion (OPE) series were not known. To obtain
more precise estimate, the radiative corrections, which are sizable,
were then computed in Refs.~\cite{Chetyrkin:2000tj,Jin:2000ek}. Unfortunately, the mass sum rule
result is highly sensitive to the parameter $s_0$ which models the
threshold of the excited states, and does not show $s_0$ stability.
Thus, Jin and K\"{o}rner conservatively estimate that the mass of the
lowest-lying $1^{-+}$ hybrid meson is larger than $1.55$~GeV \cite{Jin:2000ek}.
Nevertheless, Chetyrkin and Narison argue that $s_0$ may lie between
$3.5$ GeV$^2$ and $4.5$ GeV$^2$, so that the mass of the
lowest-lying $1^{-+}$ hybrid meson is about $1.6\sim 1.7$ GeV \cite{Chetyrkin:2000tj}.

To clarify the inconsistences in the literature,  in the present
work, we study the mass of the lowest-lying $1^{-+}$ hybrid meson using
the QCD sum rule technique. Instead of $J_\mu$, we will use
\begin{equation*}
J(x) = n^\beta n^\mu \bar d(x) \sigma_{\alpha\beta}g_s G_\mu^{\
\alpha}(x) u(x)\,.
\end{equation*}
Here and in what follows, $n^\mu$ is a light-like vector, which satisfies $n^2=0$.
So far no people adopt the above current to study the mass of the
so-called hybrid meson.  The reason that we use $J$ is because the
residue for $J$ coupling to the $1^{-+}$ hybrid meson determines the
normalization of the twist-3 light-cone distribution amplitude
(LCDA) of the lowest-lying $1^{-+}$ meson, while, for $J_\mu$, the
residue corresponds to the normalization of twist-4 LCDA. For a
relevant reaction, the amplitude related to $J_\mu$ is relatively suppressed by
$m/Q$, as compared with that due to $J$, where $Q$ corresponds
to the scale of the reaction and $m$ is the mass of the $1^{-+}$
meson. Thus $J$ may be more suitable to use to study the
lowest-lying $1^{-+}$ meson than $J_\mu$.

Note that the renormalization(RG)-improvement is always necessary in
the method of QCD sum rules because the hadronic mass is determined
by the maximum stability of the sum rule within the Borel window,
where the Borel mass is actually the renormalization scale. Unlike
the vector current $\bar q\gamma_\mu q$ which satisfies the current
conservation and is scale-independent, $J$ as well as $J_\mu$ is a
scale-dependent operator, similar to the baryonic cases
\cite{Ioffe:1981kw,Yang:1993bp}, and its anomalous dimension may
result in significant improvement in the QCD sum rule result.

The rest of this paper is organized as follows. In
Sec.~\ref{sec:qcd}, we make a brief description of LCDAs for the
$1^{-+}$ exotic meson in language of the QCD field theory.
The LCDAs play an essential r\^{o}le in the QCD
description of hard exclusive processes. Using the QCD sum rule
technique and adopting the current $J$, in Sec.~\ref{sec:sumrule} we
are devoted to the calculation of the mass sum rule of the
lowest-lying $1^{-+}$ meson.
To clarify the discrepancy between our result and those given
in the literature, we revisit the previous QCD
sum rule studies in Sec.~IV, where the current $J_\mu$ is
adopted. Finally, Sec.~\ref{sec:conclusion} contains the
conclusions and discussions.

\section{The framework of QCD}\label{sec:qcd}

Unlike the model-building approach, a $1^{-+}$ exotic meson in
language of the QCD field theory is described in terms of a set of
Fock states for which each state has the same quantum number as the
exotic meson:
\begin{eqnarray}\label{eq:fockexpansion}
|1^{-+}\rangle &=& \psi_{q\bar q} |\bar q q\rangle + \psi_{q\bar q
g} |q\bar q g\rangle + \psi_{q\bar q q\bar q} |q\bar q q \bar
q\rangle + \dots \,, \ \
\end{eqnarray}
where $\psi_i$ are distribution amplitudes and the dots denote the
higher Fock states. In contrast to the usual intuition of people,
actually a nonlocal $\psi_{q \bar q}$ does not vanish and is
antisymmetric under interchange of momentum fractions of $\bar q $
and $q$ in SU(3) limit. For instance, projecting a lowest-lying
$1^{-+}$ meson ($\pi_1$) along the light-cone ($z^2=0$), the
leading-twist LCDAs $\phi_{\parallel,\perp}$ are defined as
\begin{eqnarray}\label{eq:t-2}
\langle 0| \bar{q}_1(z) \not\!z q_2(0) |\pi_1 (P,\lambda)\rangle &=&
m_{\pi_1}^2 \epsilon^{(\lambda)}\cdot z
[\phi_\parallel], \nonumber\\
\langle 0| \bar{q}_1(z) \sigma_{\perp\nu} z^\nu q_2(0) |\pi_1
(P,\lambda)\rangle &=& i m_{\pi_1} \epsilon^{(\lambda)}_\perp p\cdot
z [\phi_\perp], \ \
\end{eqnarray}
where the notation $[\phi_{\parallel,\perp}]\equiv \int_0^1 du
e^{-i\bar upz} \phi_{\parallel,\perp} (u)$ and $p_\mu =P_\mu -z_\mu
m_{\pi_1}^2/(2pz)$ are introduced with $\bar u (u)$ being the
momentum fraction carried by $\bar q_1 (q_2)$, and the nonlocal
quark-antiquark pair, connected by the Wilson line which is not
shown, is at light-like separation. Considering the G-parity in
Eq.~(\ref{eq:t-2}), it can be known that $\phi_{\parallel,\perp}$
are antisymmetric under interchange $u \leftrightarrow \bar u$ in
SU(3) limit, i.e., the amplitudes vanish in the $z\to 0$ limit (but
{\it not} in the $z^2\to 0$ limit). This property was first
used in the study of the deep exclusive electro-production
involving a lowest-lying $1^{-+}$ exotic meson \cite{Anikin:2004vc}.
In analogy to the leading LCDAs, all twist-three two-parton LCDAs
for a $1^{-+}$ exotic meson are also antisymmetric under interchange
$u \leftrightarrow \bar u$ in SU(3) limit due to the G-parity.

$\psi_{q\bar qg}$ can be non-vanishing under interchange of momentum
fractions of quarks. Similar to the case of vector mesons
\cite{Ball:1998sk}, using non-local 3-parton gauge-invariant
operators to project amplitudes of the $|q \bar q g\rangle$, we have
three twist-3 3-parton LCDAs for a $1^{-+}$ exotic meson, built
up by a quark, an antiquark, and a gluon, where two of LCDAs are
symmetric under interchange of momentum fractions of the quark and
antiquark in the SU(3) limit, while one is antisymmetric.

\section{Evaluation of the mass of the lowest-lying $1^{-+}$
meson}\label{sec:sumrule}

Adopting the local gauge-invariant current \footnote{The other
possible choice of the twist-3 current is to consider $n^\beta
n^\alpha \bar d(x) g_s \gamma_\beta G_{\mu\alpha}(x) u(x)$. One can
also consider the 4-quark operator relevant to the $|q\bar q q\bar
q\rangle$ Fock state; however the resulting sum rule will be clouded
by the factorization of condensates.}
\begin{equation}\label{eq:current-1}
J(x) = n^\beta n^\mu \bar d(x) \sigma_{\alpha\beta}g_s G_\mu^{\
\alpha}(x) u(x)\,,
\end{equation}
we shall employ the QCD sum rules \cite{SVZ} to evaluate the mass of
the lowest-lying $1^{-+}$ exotic meson. $J(x)$ is G-parity odd,
the same as the $1^{-+}$ isovector state. The residue of $J$ coupled
to the $1^{-+}$ state is defined as
\begin{equation}\label{eq:contant-1mp}
\langle 0|J(0)|1^{-+} (p, \lambda)\rangle = f_{3,1^{-+}}^\perp
m_{1^{-+}}(\varepsilon^{(\lambda)} \cdot n)(p\cdot n)\,.
 \end{equation}
The residue constant determines the normalization of one of twist-3
3-parton LCDAs, and is also the coefficient with conformal spin 7/2
in the conformal partial wave expansion for the LCDA
\cite{Braun:2003rp}.

The method of QCD sum rule approaches the bound state problem in QCD
from the perturbative region, where non-perturbative quantities,
such as some condensates, may contribute significant corrections in
the OPE series. Through the study of
the relevant correlation function and the idea of the quark-hadron
duality, the corresponding hadronic properties, like masses, decay
constants, form factors, etc., can be thus obtained.

We consider the two-point correlation function
\begin{equation}\label{eq:correlator-1}
i\int d^4 x e^{iqx} \langle 0|T J(x) J^\dagger(0)|0 \rangle
=\Pi(q^2) (q\cdot n)^4\,.
\end{equation}
It should be noted that $J(x)$ can couple not only to $1^{-+}$ sates
but also to $0^{++}$ states as
\begin{equation}\label{eq:contant-0pp}
\langle 0|J(0)|0^{++} (p, \lambda)\rangle =-2 f_{3,S}(p\cdot n)^2\,,
 \end{equation}
where the lowest-lying state in the $0^{++}$ channel is $a_0(980)$.
Therefore, to extract the lowest-lying meson corresponding to the
$1^{-+}$ channel, at the hadron level of Eq.~(\ref{eq:correlator-1})
we shall consider two lowest-lying states. Our final result will
indicate that one of the two lowest-lying states is $a_0(980)$ and
the other one is the lowest-lying $1^{-+}$ meson, i.e., the mass of
the latter is lower than that of the $a_0(1450)$, the first excited
state in the $0^{++}$ channel. We approximate the correlation
function as
\begin{eqnarray}
\frac{4 (f_{3,a_0})^2}{m_{a_0}^2-q^2} +
 \frac{(f_{3,\pi_1}^\perp)^2}{m_{\pi_1}^2-q^2} = \frac{1}{\pi}\int^{s_0}_0 ds
\frac{{\rm Im} \Pi^{\rm OPE}}{s-q^2} \,, \label{eq:higherresonance}
\end{eqnarray}
where $\Pi^{\rm OPE}$ is the OPE result at the quark-gluon level,
and $s_0$ is the threshold of higher resonances. We apply the Borel
transformation to both sides of Eq.~(\ref{eq:higherresonance}) to
improve the convergence of the OPE series and to suppress
contributions from higher resonances. The sum rule for the first two
lowest-lying states can be written as
\begin{eqnarray}
 4\, e^{-m_{a_0}^2/M^2} f_{3,a_0}^2
 &+& e^{-m_{\pi_1}^2/M^2} (f^\perp_{3, \pi_1})^2 \nonumber\\
 &=&
\frac{1}{\pi}\int^{s_0}_0 ds\, e^{-s/M^2}\, {\rm Im} \Pi^{\rm
OPE}(s) \,, \quad \label{eq:sr-primary-1}
\end{eqnarray}
where $M$ is the so-called Borel mass.
We have checked that our $\Pi^{\rm OPE}$ result is consistent with that given in
Ref.~\cite{Ball:2006wn}, where the authors combine $\Pi^{\rm OPE}$ with another correlation function result
for calculating the twist-3 parameter of pseudoscalar mesons. Note that the authors of
Ref.~\cite{Ball:2006wn} did not perform the sum rule analysis as we
do here. To improve further the $M^2$ range of the derived QCD sum
rules, we consider the $M^2$ dependence of the various terms using
the RG equation. Thus we need to multiply each OPE term by a
coefficient $L^{(-2\gamma_J+\gamma_n)/2b}$, where $L\equiv
\alpha_s(M)/\alpha_s(\mu)$, $\gamma_J=2(7C_F/3 + N_c)$ is the anomalous dimension of $J$,
$\gamma_n$ is the anomalous dimension of the corresponding operator, and
$b=(11N_c-2n_f)/3$ with $N_c$ and $n_f$ being numbers of colors and
flavors, respectively \cite{yang:2007-1}. The anomalous dimensions
of operators appearing in the OPE series can be found in
Ref.~\cite{Yang:1993bp}. The full RG-improved sum rule thus reads
\begin{eqnarray}
 && 4\, e^{-m_{a_0}^2/M^2} [f_{3,a_0}(\mu)]^2
 + e^{-m_{\pi_1}^2/M^2} [f^\perp_{3, \pi_1}(\mu)]^2 \nonumber\\
 && =
 \Bigg\{ \frac{\alpha_s(\mu)}{360\pi^3} \int_0^{s_0}s \
 e^{-s/M^2} ds
 +\frac{89}{5184}\frac{\alpha_s(\mu)}{\pi^2}\langle \alpha_s
G^2\rangle_\mu
 \nonumber\\
& & ~~
 +\frac{\alpha_s(\mu)}{18\pi} (m_{u}\langle\bar uu\rangle_\mu
    + m_{d}\langle\bar dd\rangle_\mu) \nonumber\\
& & ~~ -\frac{\alpha_s(\mu)}{108\pi}\frac{1}{M^2}
  \big(m_{u}\langle \bar u g_s\sigma G u\rangle_\mu
 + m_{d}\langle \bar d g_s\sigma G d\rangle_\mu\big) L^{14/3b}
 \nonumber\\
&  & ~~
 +\frac{71}{729}\frac{[\alpha_s(\mu)]^2}{M^2} (\langle\bar uu\rangle_\mu^2
    + \langle\bar dd\rangle_\mu^2) L^{(b-8)/b} \nonumber\\
&  & ~~
   -\frac{32}{81}\frac{[\alpha_s(\mu)]^2}{M^2}
    \langle\bar uu \rangle_\mu \langle\bar dd\rangle_\mu L^{(b-8)/b}\Bigg\}
 L^{(b-\gamma_J)/b}\,.
 \label{eq:sr-1}
\end{eqnarray}
To subtract the contribution arising from the lowest-lying scalar
meson, we further evaluate the following non-diagonal correlation
function
\begin{equation}\label{eq:correlator-2}
i\int d^4 x e^{iqx} \langle 0|T J(x) \bar u(0)d(0)|0 \rangle
=\bar\Pi(q^2) (q\cdot n)^2\,,
\end{equation}
where the scalar current can couple to the lowest-lying scalar meson
$a_0(980)$:
\begin{equation}\label{eq:constant-2}
 \langle 0| \bar u(0)d(0)|a_0(980) \rangle
=m_{a_0} f_{a_0}\,.
\end{equation}
A similar non-diagonal correlation function was carried out in
Ref.~\cite{Ball:2006wn}, where there are additional $\gamma_5$'s in
two currents. 
Similarly, we thus obtain
\begin{eqnarray}
 && 2\, e^{-m_{a_0}^2/M^2} f_{3,a_0}(\mu) f_{a_0}(\mu)
m_{a_0}\nonumber\\
 & & ~ =  \Bigg\{ \frac{\alpha_s(\mu)}{72\pi^3} L^{-1}\int_0^{\bar s_0}s \
 e^{-s/M^2} ds
  + \frac{1}{12\pi}\langle \alpha_s G^2\rangle_\mu L^{-1}\nonumber\\
& & ~~ - \frac{\alpha_s(\mu)}{9\pi}
  (m_{u}\langle\bar uu\rangle_\mu + m_{d}\langle\bar dd\rangle_\mu)L^{-1}
  \nonumber\\
& &~~
 + \frac{2\alpha_s(\mu)}{9\pi} (m_{u}\langle\bar dd\rangle_\mu
    + m_{d}\langle\bar uu\rangle_\mu) L^{-1} \nonumber\\
 & & ~~~~~~~
   \times \bigg[ \frac{8}{3} +\gamma_E +\ln\frac{\mu^2}{M^2}
    -{\rm Ei}\bigg(-\frac{s_0}{M^2}\bigg) \bigg] \nonumber\\
 &&~~
   + \frac{1}{6M^2} \big(m_{u}\langle \bar d g_s\sigma G
d\rangle_\mu
   + m_{d}\langle \bar u g_s\sigma G u\rangle_\mu\big)L^{(14-3b)/3b}
 \nonumber\\
&  &~~
 +\frac{16}{27}\frac{\pi\alpha_s}{M^2} (\langle\bar uu\rangle_\mu^2
    + \langle\bar dd\rangle_\mu^2)L^{-8/b} \nonumber\\
&  &~~ -\frac{16}{9}\frac{\pi\alpha_s}{M^2}
    \langle\bar uu\rangle_\mu \langle\bar dd\rangle_\mu L^{-8/b}
 \Bigg\}L^{(2b-\gamma_J)/2b}
\,.
 \label{eq:sr-2}
\end{eqnarray}
Note that the RG-improvement for QCD sum rule results is very
important. For instance, $L^{(2b-\gamma_J)/2b}=1$ (or
$L^{(b-\gamma_J)/b}=1$) at $M^2=1$ GeV$^2$, while the value becomes
$0.93$ (or $1.08$) at $M^2=2$ GeV$^2$. In calculating the mass sum
rule for the lowest-lying $1^{-+}$ exotic state, in
Eq.~(\ref{eq:sr-1}) we substitute $f_{3,a_0}$ with the expression
given in Eq.~(\ref{eq:sr-2}), where use of $f_{a_0}=(0.380\pm
0.015)$~GeV and $\bar s_0=(3.0\pm0.2)$~GeV$^2$ have been made
\cite{Cheng:2005nb}. In the numerical analysis, we shall use the
following values at the scale $\mu=1$~GeV:
\begin{eqnarray}
\begin{array}{l}
  \langle \alpha_s G_{\mu\nu}^a G^{a\mu\nu} \rangle=(0.474\pm 0.120)~ {\rm GeV}^4/(4\pi)\,,  \\
  \langle \bar uu \rangle \cong \langle \bar dd \rangle =-(0.24\pm0.01)^3~{\rm GeV}^3 \,,\\
  \langle \bar ug_s \sigma G u \rangle
  \cong \langle \bar d g_s \sigma G d \rangle = (0.8\pm 0.1) \langle \bar uu
  \rangle\,, \\
   (m_u+m_d)/2=(5\pm2)\ {\rm MeV}\,.
\end{array}\label{eq:parameters}
\end{eqnarray}
Consequently, the mass sum rule for the lowest-lying $1^{-+}$
resonance can be obtained by taking the logarithm of both sides of
Eq.~(\ref{eq:sr-1}) and then applying the differential operator $M^4
\partial /\partial M^2$ to them. We choose the the Borel window
1~GeV$^2<M^2<2$~GeV$^2$, where the contribution originating from
higher resonances (and the continuum), which is defined as
\begin{eqnarray}\label{eq:res}
\frac{\frac{1}{\pi}\int_{s_0}^\infty ds\, e^{-s/M^2}\, {\rm Im} \Pi^{\rm OPE}(s)}
       {\frac{1}{\pi}\int^{\infty}_0 ds\, e^{-s/M^2}\, {\rm Im} \Pi^{\rm OPE}(s)}
 \,,
\end{eqnarray}
is less than 53\% and the highest OPE term at the quark-gluon level is no more than
13\%. Both are well under control. To obtain a reliable estimate for the mass sum rule result,
the contributions arising from higher resonances and the highest OPE term at the quark-gluon level
cannot be too large.
The excited threshold $s_0$ can be determined when the most
stable plateau of the mass sum rule result is obtained within the Borel window. Note that no
stable plateau can be obtained for the mass sum rule if the sum rule result
does not contain the RG-corrections. (See also the discussions in
Sec.~\ref{sec:sr2}.)

\begin{figure}[top]
\centering \mbox{\subfigure{\epsfig{figure=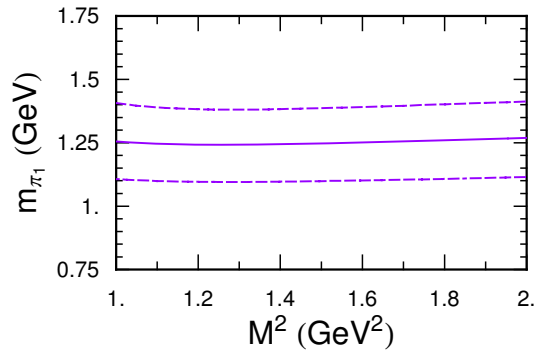,width=7cm}}}
\caption{\label{fig:sr1} The mass for the lowest-lying $1^{-+}$
exotic meson as a function of the Borel mass squared $M^2$. The
solid curve is obtained by using the central values of input
parameters. The region between two dashed lines is variation of the
mass within the allowed range of input parameters.}
\end{figure}

Numerically, we get the mass for the lowest-lying $1^{-+}$ exotic meson:
\begin{eqnarray}\label{eq:mass}
m_{\pi_1}=(1.26\pm 0.15)\ {\rm GeV},
\end{eqnarray}
corresponding to $s_0=2.5\pm 0.7$~GeV$^2$. The result for
$m_{\pi_1}$ versus $M^2$ is shown in Fig.~\ref{fig:sr1}, where the
mass is very stable within the window. From Eq.~(\ref{eq:mass}), we
know that the second lowest-lying meson for states coupling to the
operator $J$ should not be $a_0(1450)$ since the resulting mass is
lower than the experimental result of
$m_{a_0(1450)}=(1.474\pm 0.019)$~GeV \cite{PDG}. It should be stressed
that the procedure for performing the RG-improvement on the ``mass''
sum rule is very important. If the anomalous dimension of the
current $J$ was neglected, the stable sum rule could not be obtained
within the Borel window and the resulting mass was reduced by
300~MeV.

\begin{figure}[top]
\centering \mbox{\subfigure{\epsfig{figure=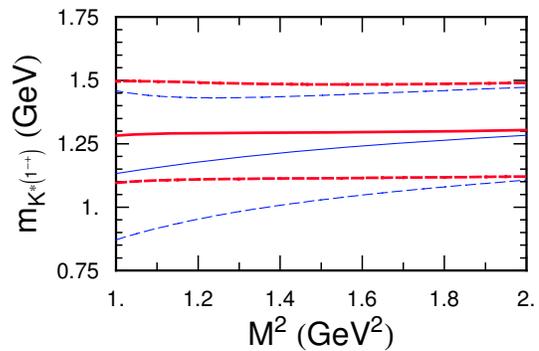,width=7cm}}}
\caption{\label{fig:sr2} The same as Fig.~\ref{fig:sr1} but for the
mass of the lowest-lying strange $1^{-+}$ exotic meson. The light (blue) and heavy (red) curves are for scenario 1 and 2, respectively, corresponding to
$s_0=2.7\pm0.7$ and $2.3\pm 0.9$~GeV$^2$.
.}
\end{figure}

We further study the existence of the $1^{-+}$ nonet. In evaluating
the mass for the lowest-lying strange $1^{-+}$ exotic meson, we use
$m_s(1~{\rm GeV})=(135\pm 15)$~MeV, $\langle \bar sg_s \sigma
Gs\rangle/\langle \bar uu\rangle \approx \langle \bar sg_s \sigma
Gs\rangle/\langle\bar uu\rangle =0.85\pm 0.05$ as additional inputs.
Because it is still questionable whether the
$\kappa(800)$ exists, we therefore consider the following two possible
scenarios. In scenario 1, the $\kappa(800)$ is treated as the
lowest-lying strange scalar meson with mass being $0.8\pm 0.1$~GeV
and $f_{\kappa}=0.37\pm0.02$~GeV, which corresponds to $\bar
s_0=2.9\pm0.2$~GeV$^2$, while in scenario 2, the $K^*_0(1430)$ is
considered as the lowest-lying strange scalar meson with
$f_{K^*_0(1430)}=0.37\pm0.02$~GeV, which corresponds to $\bar
s_0=3.6\pm0.3$~GeV$^2$. (The values are updated from
Ref.~\cite{Cheng:2005nb}.) The results are depicted in
Fig.~\ref{fig:sr2}. Using an arbitrary set of allowed inputs, within
the Borel window, the mass is stable only for scenario 2, which
hints that the $\kappa$ may not be a real particle or suitable in
the sum rule study due to its large width. Because it is not stable
in scenario 1, we assume $s_0=2.6\pm 0.7$~GeV$^2$, consistent with
the case of $\pi_1$. The result in scenario 2 is
\begin{equation}
m_{K^{*}(1^{-+})}=1.31\pm0.19\ {\rm GeV},
\end{equation}
corresponding to $s_0=2.3\pm0.9$~GeV$^2$.
%
%

\section{The mass sum rule, as derived from the correlation function given in the
literature, revisited}\label{sec:sr2}

Before concluding this paper, to clarify why our result given in the
previous section differs from those given in the literature, where a
larger mass ($\gtrsim 1.55$~GeV) for the lowest-lying $1^{-+}$ hybrid
meson was obtained, we take into account the following correction
function
\begin{eqnarray}\label{eq:correlator-3}
 && i\int d^4 x e^{iqx} \langle 0|J_\mu(x) J_\nu^\dagger(0)|0 \rangle
 \nonumber\\
 &&~~~~~
 =(q_\mu q_\nu -q^2 g_{\mu\nu})\Pi_v(q^2) + q_\mu q_\nu \Pi_s (q^2)\,,
\end{eqnarray}
with
\begin{equation}\label{eq:current-2}
J_\mu(x) = i \bar d(x) \gamma_\alpha g_s G_\mu^{\ \alpha}(x) u(x)\,,
\end{equation}
which was adopted in Refs.~\cite{Chetyrkin:2000tj,Jin:2000ek}. The
resulting sum rule can be written in the following form
\begin{eqnarray}
  e^{-m_{\pi_1}^2/M^2} m_{\pi_1}^4(f_{4, \pi_1})^2  =
\frac{1}{\pi}\int^{s_0}_0 ds\, e^{-s/M^2}\, {\rm Im} \Pi_v^{\rm
OPE}(s) \,, \nonumber\\
\label{eq:sr-primary-2}
\end{eqnarray}
where
\begin{equation}\label{eq:constant-t4}
\langle 0|J_\mu(0)|\pi_1 (p, \lambda)\rangle = f_{4,\pi_1}
m_{\pi_1}^2 \varepsilon^{(\lambda)}_\mu \,.
 \end{equation}
The $f_{4,\pi_1}$ determines the normalization of twist-4 LCDAs of
the lowest-lying $1^{-+}$ meson. The detailed OPE result for $\Pi_v$
can be found in Refs.~\cite{Chetyrkin:2000tj,Jin:2000ek}, where the
radiative corrections have been calculated.

The mass can then be obtained by applying $(M^4
\partial /\partial M^2 \ln)$ to both sides of
Eq.~(\ref{eq:sr-primary-2}). Before proceeding, three remarks are in
order.
 First, although the anomalous dimension of $J_\mu$ was
computed in Ref.~\cite{Chetyrkin:2000tj}, such a correction was not
taken into account in Refs.~\cite{Chetyrkin:2000tj,Jin:2000ek}. The
scale dependence of the operator $J_\mu$ is given by
$J_\mu(M)=J_\mu(\mu)L^{32/(9b)}$ \cite{Chetyrkin:2000tj}.
 Second, the scale-dependence of the strong-coupling constant is ignored in
Refs.~\cite{Chetyrkin:2000tj,Jin:2000ek}.
 Third, for theoretical completeness, the RG effect should be included
in the mass sum rule (\ref{eq:sr-primary-2}) as done in
Eq.~(\ref{eq:sr-1}), so that the right hand side of
Eq.~(\ref{eq:sr-primary-2}) should be multiplied by the overall
factor $L^{1-64/(9b)}$. Such a factor equal to $1$ at $M^2=1$ GeV$^2$,
but becomes $\sim 0.95$ at $M^2=2$ GeV$^2$.

\begin{figure}[b]
\centering \mbox{\subfigure{\epsfig{figure=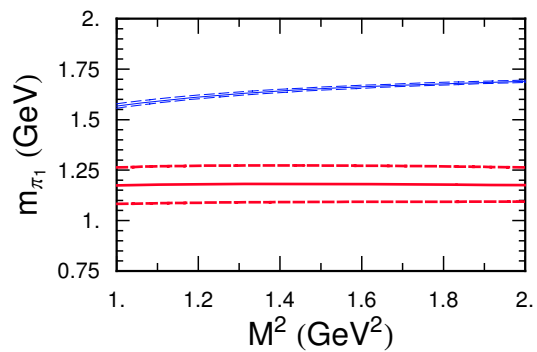,width=7cm}}}
\caption{\label{fig:sr3} The same as Fig.~\ref{fig:sr1} but from the
sum rule given in Eq.~(\ref{eq:sr-primary-2}). The heavy (red)
curves are obtained from RG-improved sum rule, where the solid curve corresponds
to $s_0=2.4$ GeV$^2$. The light (blue)
curves, using $s_0=4.0$ GeV$^2$ and the scale $\mu=2$ GeV for $\alpha_s$ and remaining parameters,
do not contain RG corrections.}
\end{figure}

In Fig.~\ref{fig:sr3}, we plot the mass sum rule as a function of
the Borel mass squared within 1~GeV$^2<M^2<2$~ GeV$^2$. The light curves
(the upper set) are the results,\footnote{In Ref.~\cite{Chetyrkin:2000tj}, Chetyrkin and Narison use the finite energy sum rule approach
in the analysis and argue that $s_0$ may lie between $3.5$ GeV$^2$ and $4.5$ GeV$^2$. They obtain the mass of the
lowest-lying $1^{-+}$ hybrid meson to be $1.6\sim 1.7$ GeV. In their analysis, the used parameter $1/\tau$
roughly agrees with $M^2\approx (1.6\sim 2)$~GeV$^2$. We thus find that the light curves in Fig.~\ref{fig:sr3}
are consistent with the result obtained by Chetyrkin and Narison.}
where we simply fix the scale at 2 GeV for
input parameters and use $s_0=4$~GeV$^2$, which consist with that obtained by Jin
and K\"{o}rner \cite{Jin:2000ek} (see Fig.~3 in Ref.~\cite{Jin:2000ek}).
Moreover, if RG effects are not considered, as in Ref.~\cite{Jin:2000ek},
there is no plateau for any value of $s_0$. Nevertheless, the stable
plateau can be reached for a typical value of $s_0$ if we take into account
the RG corrections in the sum rule.
See the heavy curves (the lower set) in Fig.~\ref{fig:sr3}. The corresponding results are
\begin{eqnarray}\label{eq:mass-2}
m_{\pi_1}&=&(1.18\pm 0.09)\ {\rm GeV},\nonumber\\
s_0&=& (2.4^{+0.4}_{-0.2}) \ {\rm GeV}^2\,,
\end{eqnarray}
where the errors are due to variation of the input parameters.
Note that the plateau covers the range for $M^2\sim 1.5$~GeV$^2$. Unfortunately,
for 1~GeV$^2<M^2<2$~ GeV$^2$ and $s_0=2.4$~GeV$^2$, the
contribution originating from higher resonances (and the continuum) lies between 72\% and 95\%,
which is too large, so that the resulting mass is less reliable.
%


\section{Conclusions and discussions}\label{sec:conclusion}

In this work we have studied the mass of the lowest-lying $1^{-+}$ hybrid meson using
the QCD sum rule technique. Instead of $J_\mu(x)= i \bar d(x)
\gamma_\alpha g_s G_\mu^{\ \alpha}(x) u(x)$ which is always adopted in the literature,
we use, for the first time,
\begin{equation*}
J(x) = z^\beta z^\mu \bar d(x) \sigma_{\alpha\beta}g_s G_\mu^{\
\alpha}(x) u(x)\,,
\end{equation*}
as the interpolating current of the lowest-lying $1^{-+}$ hybrid meson.
 Our main results are as follows:
\begin{itemize}
\item Using $J$ as the interpolating current of the $1^{-+}$ exotic mesons, which are the
so-called $1^{-+}$ hybrid mesons (see the discussions in Sec.~\ref{sec:qcd}), we have obtained the mass
for the lowest-lying $1^{-+}$ exotic meson to be $(1.26\pm 0.15)$~GeV. The threshold of excited states, $s_0$,
as well as the resulting mass can be thus determined
due to the existence of the most stable plateau of the mass sum rule result within the Borel window,
where the contributions arising from the resonances and the highest OPE term at the quark-gluon level
are well under control.
The plateau for the mass result versus the Borel mass squared exists only when the RG-corrections are
taken into account in the sum rule. Our result is in good agreement with the world
average of data for the mass of $\pi_1(1400)$, which is $m_{\pi_1(1400)}=(1.376\pm 0.019)$ GeV.
\item On the other hand, to clarify the discrepancy between our result and those given in the literature
for which $J_\mu$ is used as the interpolating field of the $1^{-+}$ hybrid meson, we have reexamined the previous
sum rule calculation. (See Sec.~\ref{sec:sr2}.) As in Ref.~\cite{Jin:2000ek},
the curve for the resulting mass versus the Borel mass squared has no plateau region for any value of $s_0$
if RG corrections are not considered. Unfortunately, with RG-corrections,
the plateau covers the range for $M^2\sim 1.5$~GeV$^2$, where the
contribution arising from higher resonances is $\gtrsim 80\%$ in the sum rule result,
so that the resulting mass is less reliable.

\item Using $J$ in the sum rule study, we also obtain the mass for the strange $1^{-+}$ hybrid meson:
$m_{K^*(1^{-+})}=1.31 \pm 0.19$ GeV. In the Particle Data Book (PDG) \cite{PDG}, the two
strange mesons, $K^*(1410)$ and $K^*(1680)$, are currently assigned
to be $2^3S_1$ and $1^3D_1$ states, respectively. However, because
the $K^*(1410)$ is too light as compared with the remaining $2^3S_1$
nonet states, therefore it could be replaced by $K^*(1680)$ as the
$2^3S_1$ state. If so, our result hints that the $K^*(1410)$ is very
likely to belong to the lowest-lying $1^{-+}$ nonet.
\item We have presented a discussion for properties of the
$1^{-+}$ exotic meson (which is usually called the $1^{-+}$ hybrid meson)
based on the QCD field theory. Although  $q$-$\bar q$-$g$ Fock states of the exotic meson are suppressed (see also discussions in Sec. I),
the twist-2 distribution amplitudes, which are
antisymmetric under interchange of the quark and anti-quark momentum fractions in the SU(3) limit,
give main contributions to the deep exclusive electro-production involving a lowest-lying $1^{-+}$ exotic meson. Thus, from the point of view of QCD, the cross section of the above reaction is still sizable and the exotic (hybrid) meson can be studied in experiments at JLAB, HERMES or Compass \cite{Anikin:2004vc}. Furthermore, because of the non-small first Gegenbauer moment of the twist-2 distribution amplitudes, the branching ratios for two-body $B$ decays involving the $\pi_1(1400)$ \cite{kcy-Bpi1} could be of order $10^{-6}$, which are easily accessible at the $B$ factories.

\end{itemize}

In conclusion,  we may have evidence for the
existence of $1^{-+}$ exotic mesons. These states are allowed in Quantum Chromodynamics
but not in the conventional quark model. Theorists tried to predict these states, especially with mass $\lesssim$ 1.6
GeV, but did not succeed. In this paper, we have shown that the $\pi_1(1400)$
state is indeed expected in QCD. We also predict the mass of the lowest-lying strange
$1^{-+}$ exotic meson which is a little larger than $m_{\pi_1 (1400)}$. Thus, the observation of the $\pi_1(1400)$ state
can be the direct evidence of QCD in explaining the confinement mechanism of
strong interactions.


\acknowledgments

\vskip-4.5mm  This work was supported in part by the National Science Council of
R.O.C. under Grant No: NSC95-2112-M-033-001.



\end{document}